\documentclass[usenatbib]{mnras}

\usepackage{newtxtext,newtxmath}

\usepackage[T1]{fontenc}
\usepackage{ae,aecompl}

\usepackage{graphicx}	
\usepackage{amsmath}	
\usepackage{amssymb}	
\usepackage{longtable}
\usepackage{CJKutf8}

\newcommand{\E}[1]{\times 10^{#1}}

\newcommand{\degree}{\ensuremath{^\circ}}

\title[ASKAP Single Pulse Galactic Plane Survey]{A survey of the Galactic plane for dispersed radio pulses with the Australian Square Kilometre Array Pathfinder}

\author[H. Qiu et al.]
{Hao Qiu(\begin{CJK*}{UTF8}{gbsn}邱昊\end{CJK*})$^{1,2}$\thanks{E-mail:hqiu0129@uni.sydney.edu.au},
K. W. Bannister$^{2}$,
R. M. Shannon$^{3,4}$,
Tara Murphy$^{1}$,
\newauthor{Shivani Bhandari$^{2}$, Devansh Agarwal$^{5,6}$, D. R. Lorimer$^{5,6}$ and
J. D. Bunton$^{2}$}
\\
$^{1}$Sydney Institute for Astrophysics, School of Physics, University of Sydney, Sydney, New South Wales 2006, Australia\\
$^{2}$CSIRO Astronomy and Space Science, Australia Telescope National Facility, P.O. Box 76, Epping, NSW 1710, Australia\\
$^{3}$Centre for Astrophysics and Supercomputing, Swinburne University of Technology, Mail H30, PO Box 218, Hawthorn, VIC 3122, Australia\\
$^{4}$ARC Centre of Excellence for Gravitational Wave Discovery (OzGrav)\\
$^{5}${West Virginia University, Department of Physics and Astronomy, P. O. Box 6315, Morgantown, WV, USA}\\
$^{6}${Center for Gravitational Waves and Cosmology, West Virginia University, Chestnut Ridge Research Building, Morgantown, WV, USA}\\
}

\date{Accepted 2019 March 08. Received 2019 January 31; in original form 2018 November 30}

\pubyear{2018}

\begin{document}
\label{firstpage}
\pagerange{\pageref{firstpage}--\pageref{lastpage}}
\maketitle

\begin{abstract}
We report the results from a survey of the Galactic plane 
for dispersed single pulses using the 
Australian SKA Pathfinder (ASKAP).
We searched for
rare bright dispersed radio pulses comprising 160 pointings covering 
4800 deg$^2$  of the Galactic plane within ${|\text{b}|< 7 \degree}$,
each pointing with an exposure time of 10 hours.
We detected one fast radio burst, FRB 180430, 
and single pulses from 11 pulsars. 
No rotating radio transients were detected. We detected
FRB 180430 in the Galactic plane in the anticentre direction
with a fluence of 216$\pm5$~Jy~ms a dispersion measure (DM) of 
264.1 $ \text{pc} \ \text{cm}^{-3}$.
We estimate the extragalactic DM of the object to be less than 86.7 $ \text{pc} \ \text{cm}^{-3} $ depending on the electron density model.
One model suggests
that this FRB may be a giant pulse within our galaxy; we discuss how this may not correctly represent the line-of-sight DM.
Based on the single detection of FRB 180430 in $3.47 \E{4}\deg^2 \text{h}$ we derive
a FRB event rate in the 
Galactic plane
at the 20 Jy~ms threshold to be
in the range 2--140 per sky per day
at 95\% confidence. Despite the necessarily
large uncertainties from this single detection, this is consistent with the current ASKAP 
all-sky detection rate.
\end{abstract}




\section{Introduction}

Single pulse radio transients such as fast radio bursts \citep[FRBs;][]{2007Sci...318..777L}, pulsar giant pulses \citep{1968Sci...162.1481S} and 
rotating radio transients \citep[RRATs;][]{2006Natur.439..817M} are isolated dispersed bursts of 
radio emission occurring on an extremely short timescale of nanoseconds to milliseconds.
FRBs and single pulses of RRATs are distinguished by the dispersion measurement (DM) of these pulses. The DM of FRB are higher than the Milky Way DM contribution in that direction while the DM of pulsars and RRATs are lower than the Galactic contribution. Most RRATs have been observed with
recurring outbursts but no FRB, except FRB 121102, has been observed to repeat.

The nature of FRBs remain largely unknown 
due to lack of definitive observational data.
The DM of FRBs and RRATs indicate a 
different distance scale for the two populations.
However, the sporadic burst rate and pulse properties of transient radio neutron stars such 
as RRATs (\citealp{2011BASI...39..333K} \& \citealp{2009MNRAS.400.1431M}) and magnetars \citep{2015ApJ...807..179P} display 
a similarity to the sporadic bursts of the repeating 
FRB 121102 \citep{2014ApJ...790..101S,2018ApJ...863....2G}.

The Galactic plane is an interesting region to search for FRBs and single pulses from RRATs. 
The population of RRATs is observed to have a Galactic
distribution\footnote{http://astro.phys.wvu.edu/rratalog/}, with 69 out of the total 113 in low/intermediate Galactic latitudes. 
For FRBs, 11 
low/intermediate latitude (${|\text{b}|< 15 \degree}$) FRBs have been
detected from a total of 66 \citep{2016PASA...33...45P}\footnote{http://frbcat.org/}.
The paucity of FRBs detected in the Galactic plane is because the large majority of these FRB searches having been 
performed simultaneously with pulsar searches
which
focus on lower Galactic latitudes.

The search for FRBs in the southern Galactic plane
has benefited from 
several large scale radio survey projects:
the Parkes Multibeam Pulsar 
Survey \citep[PMPS;][]{2001MNRAS.328...17M},
the High Time Resolution Universe South \citep[HTRU--S;][]{2010MNRAS.409..619K},
the upgrade to the Molonglo Observatory 
Synthesis Telescope \citep[UTMOST;][]{2017PASA...34...45B},
and the SUrvey for Pulsars and 
Extragalactic Radio Bursts \citep[SUPERB;][] {2018MNRAS.475.1427B}. 
These radio transient and pulsar
 surveys by Parkes (PMPS, HTRU--S, SUPERB)
have provided 30 minutes to 1 hour pointings of 
the Galactic plane.
These surveys have good sensitivity but have a relatively lower 
integration time per pointing due to the limited field of view (FoV) of these telescopes. 
With a telescope that has much larger FoV, it is possible to cover more sky simultaneously and provide longer integration time per pointing with the same amount of observation time.
Thus increase the probability of detecting more of these bright dispersed single pulses, rarer RRAT outbursts and repetition.

The Australian SKA Pathfinder (ASKAP; 
\citealp{2008ExA....22..151J}; 
\citealp{2012SPIE.8444E..2AS}) is a 
new wide-field radio telescope designed for large surveys. 
ASKAP has 36 12-metre antenna dishes which are equipped 
with an approximate 30 deg$^2$ FoV phased array feed receivers
\citep[PAF;][]{2008RaSc...43.6S04H}. 
Each single ASKAP antenna has an measured System Equivalent Flux Density of approximately 2000 Jy, 
which is less sensitive than the Parkes radio telescope by a factor of 50,
but the 30 deg$^2$ field of view at 1.4 GHz provided by the PAF is 40 times wider than Parkes.
This enables a survey with low sensitivity but large simultaneous coverage.

The Commensal Real-time ASKAP Fast Transients 
survey \citep[CRAFT;][]{2010PASA...27..272M}, has developed high time resolution ($\sim$1 ms)  capabilities on ASKAP with the goal of localising dispersed radio pulses.
22 bright FRBs (fluences $> 20$ Jy\,ms) have been reported from surveys at Galactic latitude 50$\degree$ 
\citep{2017ApJ...841L..12B,2018Natur.562..386S,2018arXiv181004353M} 
and two at a Galactic latitude of 20$\degree$ \citep{2018arXiv181004353M}.

In the past works, the discovery rates of FRBs
at low Galactic latitudes is lower compared 
to high latitudes (\citealp{2014ApJ...792...19B}, \citealp{2014ApJ...789L..26P}
and \citealp{2015MNRAS.447.2852K}).  
Several factors have been proposed to explain this absence by \citet{2014ApJ...789L..26P}: 
dispersion and scattering by the interstellar medium (ISM); relatively high sky temperature and scintillation.
There is also some evidence that suggest at least one low Galactic latitude FRB may be RRATs 
due to the additional H$\alpha$ and H$\beta$ observations for 
FRB 010621\citep{2014MNRAS.440..353B}. 
This implies a possible population of giant pulses from distant pulsars that were not previously detected similar to FRBs.
 However, recent FRB discoveries such as \citet{2017MNRAS.468.3746C}
have shown less dependence on Galactic latitude in detectability.

To search for  low burst rate (> 1 $\text{Hrs}^{-1}$) RRATs and constrain the low-latitude FRB event rate,
we conducted a millisecond timescale Galactic plane survey for pulsars and radio transients 
 with the longest observation time per pointing,  
aimed at 10-hours total observation time of the entire Galactic plane observable from ASKAP.


\section{Survey Overview}
\subsection{Receiver Configuration and Survey Data Reduction Pipeline}
The PAF consists of 36 dual polarization beams sensitive to
radio frequencies between 0.7 and 1.8 GHz.
Beam weights are calibrated with a maximum signal-to-noise (S/N) algorithm and observing 
the sun as the reference source (\citealp{2014PASA...31...41H}; \citealp{2016PASA...33...42M}).
The signal received by the PAF from each element is digitized and channelized to 336 1-MHz frequency channels. 
For the observations in this survey the band was centred at 1297 MHz.

The CRAFT data pipeline is described in detail by \citet{2014JAI.....350004C}. 
For the survey presented in this work,
the voltages from the beamformer are squared and averaged over an integration time of 1.265 ms. 
The data is then transmitted to a processing computer and saved in \textsc{sigproc} format filterbanks  \citep{2011ascl.soft07016L}. 

The data are then searched offline with the Fast Real-time Engine for Dedispersing Amplitudes (FREDDA; Bannister et al. in prep.), 
a GPU based FRB detection pipeline implementing the Fast Dispersion Measure Transform algorithm \citep[FDMT;][]{2017ApJ...835...11Z}. There is a detailed description of the software in \citet{2017ApJ...841L..12B} 

\begin{figure}
\centering
	\includegraphics[width=\columnwidth]{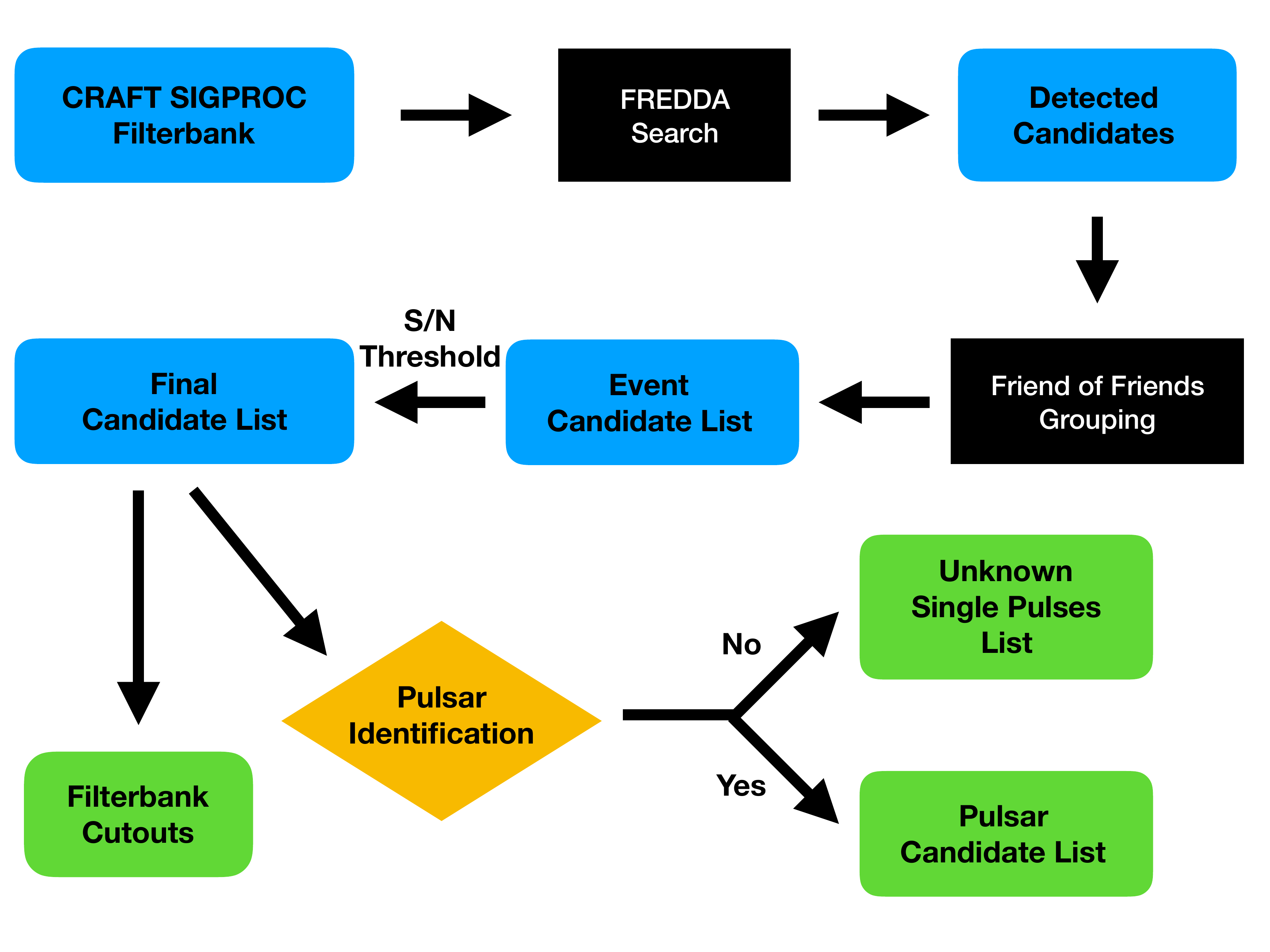}
    \caption{Candidate Selection Pipeline Flowchart: The blue boxes indicate the CRAFT pipeline products, 
    the black boxes represent the FREDDA detection and filtering processes,
    the yellow diamond is the post processing analysis and the green boxes show the final products of the pipeline.}
    \label{fig:pipeline}
\end{figure}

In this survey we use FREDDA to search for candidates within the dedispersion range is $0-3763\ \text{pc} \ \text{cm}^{-3}$.
 To focus our search on possible FRB candidates which have a relative narrower pulse profile and reduce the misidentification of known radio frequency interference (RFI),
we applied a width boxcar threshold of $< 11$ time samples (approximately 13.915 ms)
and a S/N $>7.5$ detection limit to reduce false candidate events.
Due to the significant number of pulsars in the Galactic plane,
we performed a pulsar identification using the beam position
and the DM of the candidates.
We searched for pulsars and RRATs within a 1.5 beam radius (approximately 0.68 degrees) and 10 percent DM threshold based on DM and position information from 
the latest version of the pulsar catalogue
{\textsc{psrcat}\footnote{http://www.atnf.csiro.au/research/pulsar/psrcat/} database 
\citep{2005AJ....129.1993M}. 
The pulsar and RRAT candidates that match one of the following criteria were selected: 
\begin{enumerate}
\item The object has been detected more than 5 times in all observations during this survey. 
\item The object has been detected less than 5 times but has an average significance of $9.5 \sigma$ or higher during this survey.
\end{enumerate}
Identified single pulses and unidentified single pulses are written to separate files
for further RFI check and pulse validation.
For strong candidates (S/N $> 9.5$), five second cutouts were automatically generated to display the spectra of the filterbank for fast visual validation check of RFI.
We then visually inspected the cutout images for false detections.
We also show a visualization of this pipeline in Figure \ref{fig:pipeline}.}

The amplitude of candidate spectra is
converted to physical units with an assumed system 
equivalent flux density (SEFD) measurement ($S_{sys}=2000\ \text{Jy}$).  
We assume off-pulse noise in the observation data to be gaussian noise.

\subsection{Survey Parameter Space}
\begin{figure}
    \centering
	\includegraphics[trim={0 0 0 0 },clip,width=\columnwidth]{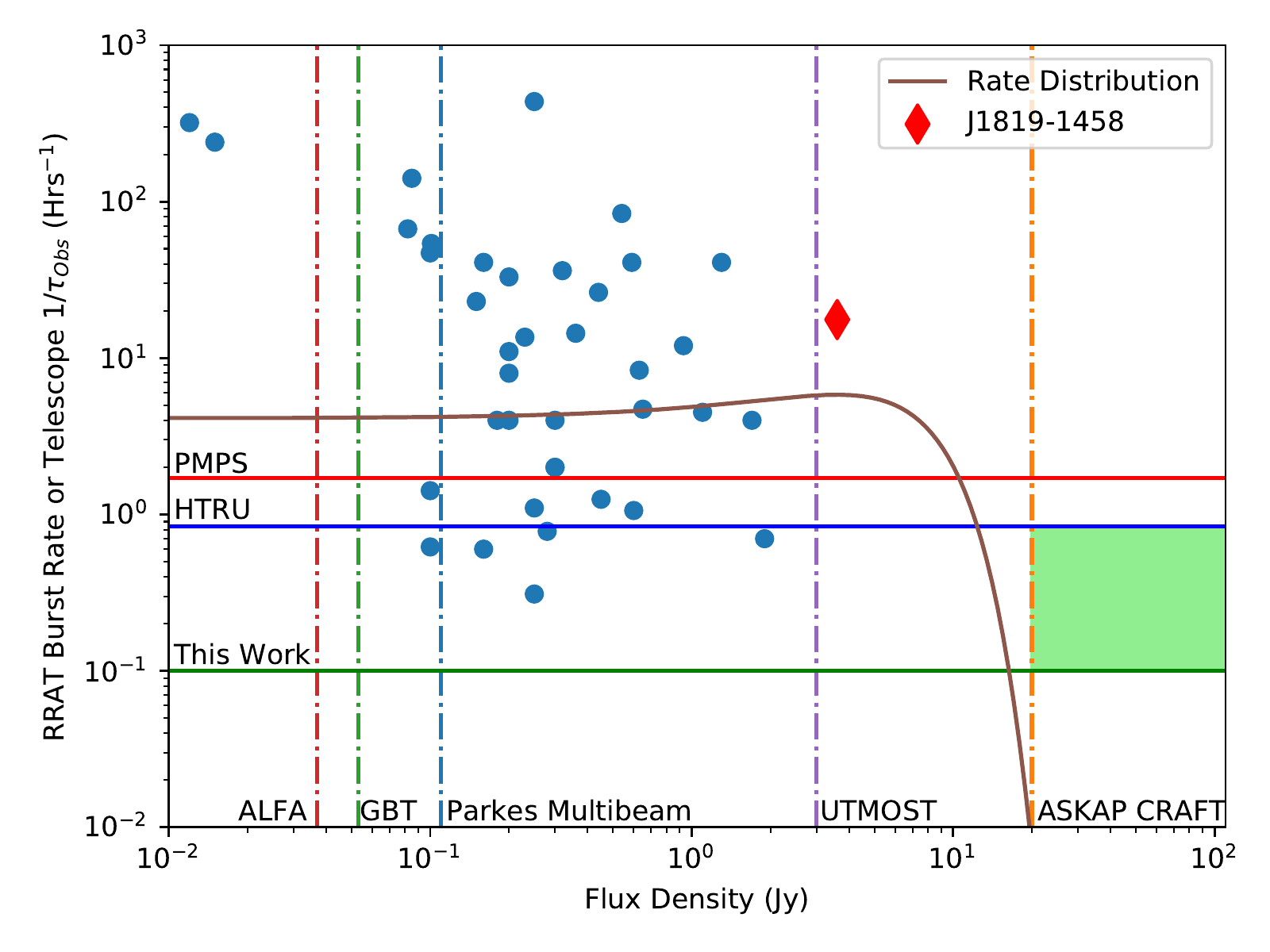}
    \caption{Comparison of different surveys: The estimated single pulse sensitivity threshold
    limits for ASKAP-CRAFT, UTMOST, the Parkes Multibeam Receiver, Green Bank Telescope (GBT) and the Arecibo L-band Array Feed (ALFA) as vertical dashed lines. 
    The total integration time
    $\tau_{\rm obs}$ for 
    previous Parkes surveys (PMPS and HTRU-S) and 
    our Galactic Plane survey in this work with horizontal lines. 
    The average flux density and burst rate 
    of known RRATs (blue dots) recorded 
    in the RRAT catalog 
    are plotted for comparison. 
    We highlight the brightest RRAT J1819-1458 (red diamond) 
    and plot an estimated rate distribution for 
    bursts of different flux densities. The new parameter space we will probe is show in the light green region.}
    \label{fig:param_space}
\end{figure}

We conduct a search 
for bright single pulses along the Galactic plane in this survey 
using 8 ASKAP antennas as 8 separate single dish telescopes. 
With the extended 240 deg$^2$ FoV of 8 ASKAP antennas this is an ideal survey for capturing random 
rare bright radio bursts such as FRBs and pulsar giant pulses. 
This survey has a significantly poorer sensitivity compared 
with other single dish telescopes such as Parkes but surpasses 
the integration time per pointing of most other Galactic plane surveys while only using a much shorter observing period as 
shown in Figure \ref{fig:param_space}.
We highlight the light green region as the previously undetected time domain for
burst repetition that we probed with this survey.
Also, the single dish sensitivity of ASKAP is sufficient for detecting previously detected FRBs, 
and we expect to search for the population of bright FRBs as displayed in \citet{2018Natur.562..386S} at higher Galactic latitudes.

In Figure~\ref{fig:param_space},
we plot 40 RRATs with known burst rates and flux measurements to compare 
the current population of RRATs to 
the sensitivity and observation time of this previous Galactic plane surveys. 
We also display the single pulse distribution 
of the RRAT J1819-1458 based on
the work of single pulse energetics in \citet{2012MNRAS.423.1351B}.
This limit indicates that this survey will not likely detect the fainter frequent Galactic radio single pulses from these known RRATs. 
It should be noted that there are RRATs with lower flux densities than J1819-1458 
but without a known burst rate we do not estimate
the detectability of these RRATs in this plot.
To confine our estimates, we calculate an estimate detection rate with a simulated population of RRATs in Section 2.4. 

\subsection{Survey Strategy and Observations}
This survey operates ASKAP in `fly's-eye' mode, pointing each dish in a different direction to
extend the total FoV during observing.
The FoV in `fly's-eye' observing mode 
is the beam pattern using a single ASKAP dish, 
but it is multiplied by the number of dishes we deploy for observing. 

For the CRAFT survey, we have adopted a beam pattern configuration as shown by Supplementary Figure 2 in
\citet{2018Natur.562..386S}.
Eight ASKAP antennas were 
operated in fly's-eye mode during this survey,
giving a maximum simultaneous
coverage of $36\times8$ $0.9 \degree$ circular beams on the sky.
\begin{figure}
\centering
 	\includegraphics[trim={0 0.1cm 0 0.1cm },clip,width=\columnwidth]{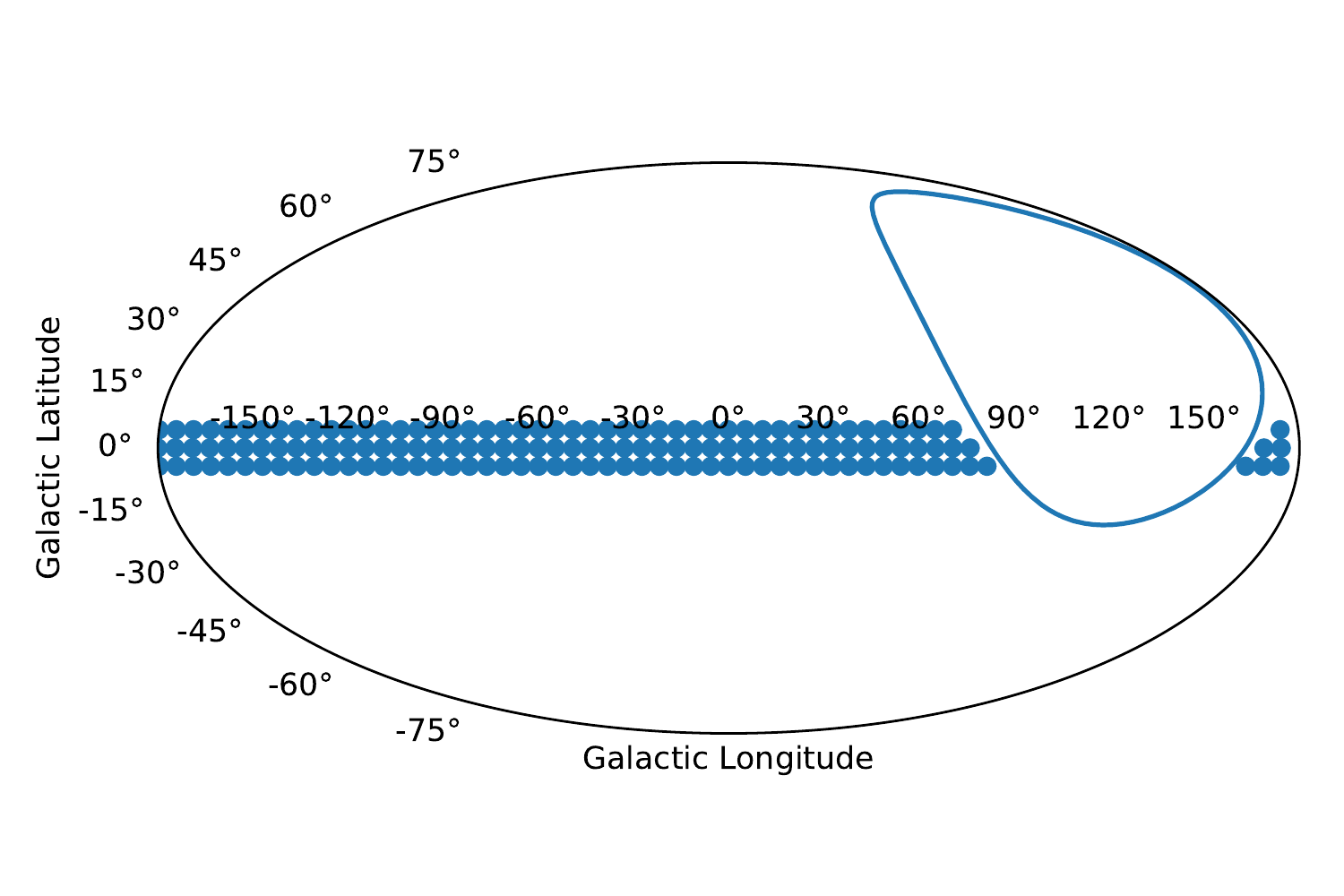}
    \vspace{-10mm}
    \caption{Pointings along the Galactic plane are marked in blue dots, the blue line indicates declination limit of the telescope.}
    \label{fig:gpfp}
\end{figure}

\begin{figure}
\centering
	\includegraphics[trim={0 0.0cm 0 0.0cm },clip,width=\columnwidth]{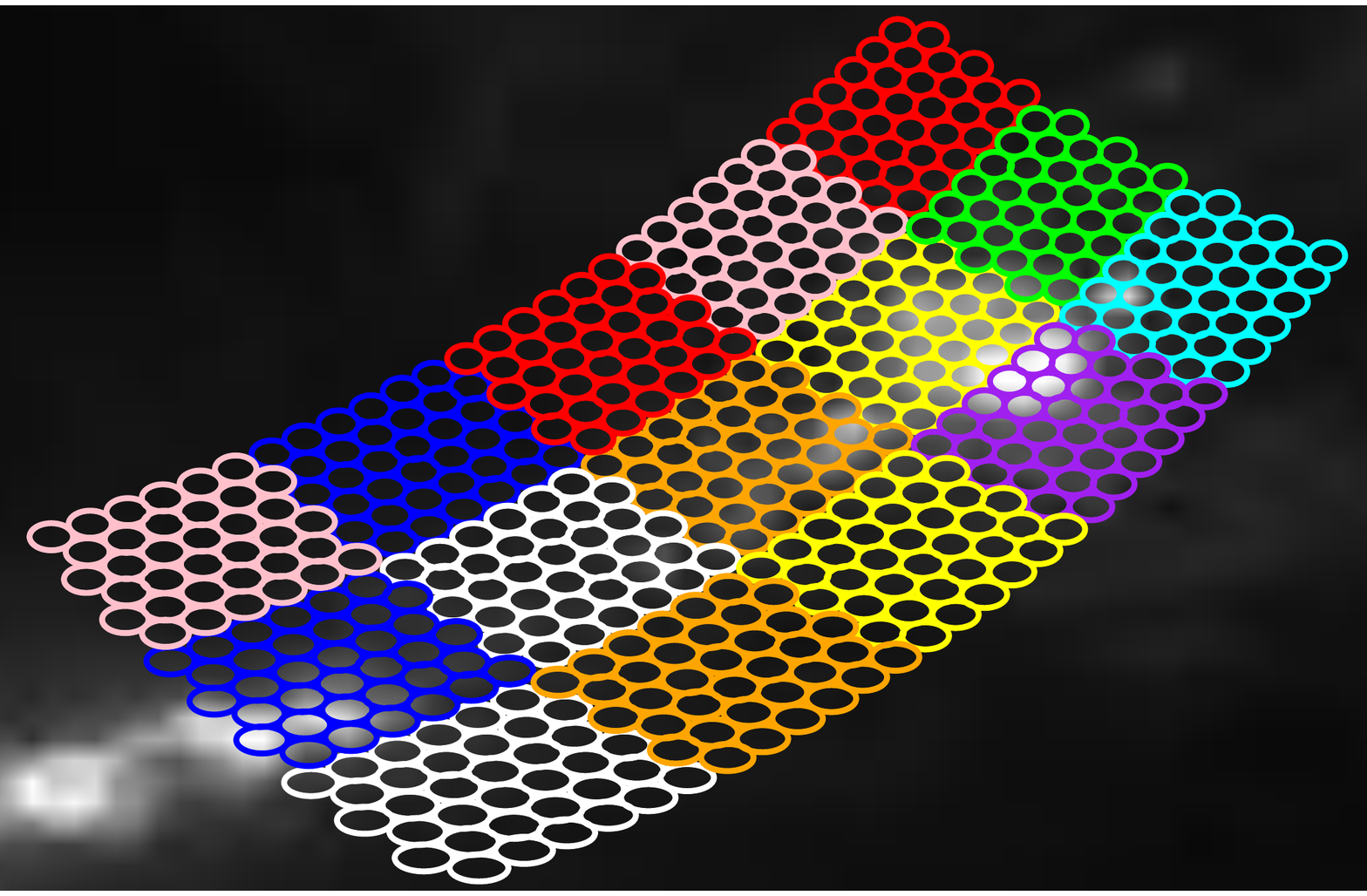}
    \vspace{-20mm}
    \caption{An example of footprint patterns of observed pointings meshed on sky along the galactic plane, we use a random colour scheme to distinguish the footprints.
    }
   \label{fig:rainbow}
\end{figure}

Our survey is designed to observe within the Galactic latitude range $|\text{b}| <7 \degree$, we display the observable region in Figure \ref{fig:gpfp}.
Each ASKAP antenna has a roll axis for the prime reflector to correct apparent rotation of the sky.
160 pointings for the survey were planned by meshing the PAF beam pattern footprint. We align the antenna footprint into three parallel strips of observation footprints perpendicular to the Galactic pole at respectively b~$= 0\degree, \pm4.67 \degree$ 
as shown in Figure \ref{fig:rainbow}
. This meshes the footprints together perfectly for maximum survey efficiency.

Observations for this survey were carried out between 2018 March 29th 
and 2018 May 30th.
Each pointing region was to be observed with a total time of 10 hrs, 
this sums to a total observation time of 66.7 antenna days. 

\begin{table}
	\centering
	\caption{Survey Observation Specifications}
	\label{tab:obs_spec}
	\begin{tabular}{lc} 
		\hline
        Region & $|b|< 7\degree$, $170\degree<l<180\degree$,  \\ &$-180\degree<l<75\degree$\\
		 $\tau_{obs,pointing}$ (s) & 36\,000 \\
	SEFD (Jy)	&  $\sim$2000 \\
	Total pointings & 160 \\
	$\tau_{obs,total}$ (s) & 5\,760\,000 \\
        $N_{beams}$ per pointing & 36\\
        FWHM 1.32 GHz & $0.9 \degree$ \\
		Bandwidth (MHz) & 336 \\
        $\tau_{samp}$ (ms)& 1.265\\
		$\Delta \nu_{chan}$ (MHz) & 1\\
        Centre Frequency (GHz)& 1.297\\
		$N_{chans}$ & 336\\

		\hline
	\end{tabular}
\end{table}
We show the technical specifications for the observation in Table \ref{tab:obs_spec}.
Observation time for some pointings 
did not reach full target time due to commissioning work during the observation period.
In this paper, we present results obtained from a total observation time of 63.1 antenna days.

\subsection{RRAT Population Model}
\begin{figure}
    \centering
	\includegraphics[trim={0 0 0 0 },clip,width=\columnwidth]
    {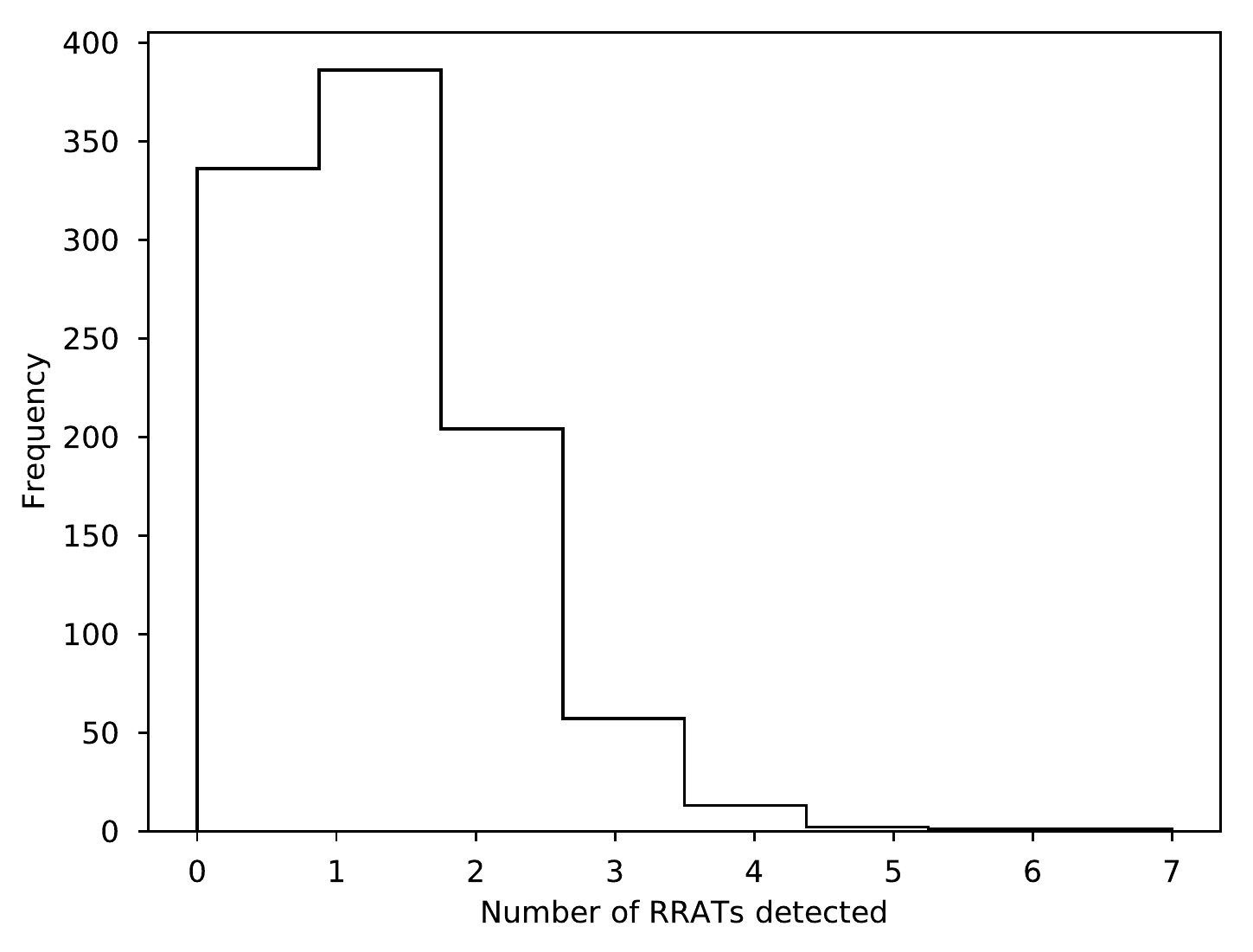}
    \vspace{-8mm}
    \caption{Predicted detection of RRATs using parameters from this survey with a simulated RRAT population.}
    \label{fig:rrat_sim}
\end{figure}

In order to get a prediction of the number of RRATs detectable by our survey, we use the population model as described in  Agarwal et al. (in prep). 
Using \textsc{PsrPopPy2}\footnote{https://github.com/devanshkv/PsrPopPy2} \citep{b8s2014} we create a simple population model without any time evolution. 
In brief, we start with a uniform distributions in spin period, $P$, 
Galactic scale height, $z$, Galactocentric radius, $R$, 
and log-uniform in pseudo-luminosity, $L$, and burst rate, 
$\dot{\chi}$. For each RRAT, the single pulse amplitude distribution 
is taken to be log-normal with mean sampled from the luminosity 
distribution and standard deviation set to ten percent of the mean. 
The population is then passed through four model surveys 
: the Parkes multibeam survey \citep{2001MNRAS.328...17M,k11}, 
two higher latitude surveys  \citep{Jacoby, Edwards}, 
and the  HTRU intermediate latitude survey \citep{2010MNRAS.409..619K}. 
The distributions of the detected population compared with observed 
distributions form the observed RRATs. 
The underlying population is then modified using correction factors in a similar manner to that described in \cite{Lorimer2006}. 
We reach the optimal underlying population when the reduced $\chi^2$ between the distributions of the model detected population and observed population 
are of order unity.  
We draw 54 RRATs (as detected by the above mentioned four surveys) using the optimal underlying distributions for the physical parameters and run a model CRAFTs survey to get the number of potentially detectable RRATs. 
To produce the distribution of the number of RRATs detectable by the CRAFT survey,
we run a 1000 iterations of the following simulation, shown in Figure \ref{fig:rrat_sim}.


\section{Results}

\subsection{Pulsar and RRAT detections}
\begin{table*}
	\centering
	\caption{List of confirmed pulsar detections during the survey}
	\label{tab:gp}
	\begin{tabular}{lrrrrrr} 
		\hline
		Pulsar & Number  of& Average S/N & DM  &
        Boxcar Width$^a$  & Flux 1.4GHz $^b$ &
        Pulse Width$^b$
         \\
  		& Detections&  & ($ \text{pc} \ \text{cm}^{-3}$) & (samples)
         & (mJy) &
        (ms)       
         \\
        \hline

B0525+21 & 11 & 11.0 & 52.51 & 7.00 & 9.00 & 214.2 \\
B0531+21 & 746 & 13.8 & 56.56 & 0.39 & 14.00 & 4.7 \\

B0833--45 & 195013 & 14.2 & 67.40 & 1.17 & 1100.00 & 4.5 \\
B0835--41 & 808 & 10.5 & 147.1 & 3.15 & 16.00 & 18.0 \\

B1641--45 & 72394 & 12.2 & 477.7 & 7.27 & 296.40 & * \\

B1727--47 & 42 & 11.5 & 122.4 & 2.98 & 12.00 & 32.0 \\
B1749--28 & 26 & 9.40 & 49.70 & 4.15 & 18.00 & 15.0 \\

B1933+16 & 32 & 9.94 & 159.0 & 4.19 & 42.00 & 17.7 \\

B2020+28 & 129 & 10.3 & 24.69 & 1.11 & 38.00 & 15.8 \\

J1047--6709 & 2 & 9.97 & 115.9 & 1.00 & 4.00 & 21.0 \\
J1107--5907 & 11 & 13.6 & 40.75 & 5.45 & 0.18 & 170.0 \\
\hline
		\hline
		\multicolumn{3}{l}{a. The boxcar width is in units of 1.2 ms samples.}\\
\multicolumn{3}{l}{b. Literature results provided from \textsc{psrcat}}
	\end{tabular}
\end{table*}

A total of 11 pulsars were detected through single pulse detection in this survey 
as shown in Table \ref{tab:gp}.  
We did not detect any RRATs listed in \textsc{psrcat} or the \textsc{rratalog}.
The period and flux density values recorded in \textsc{psrcat} are also displayed for comparison. 
We note that the mean flux density from literature is the flux averaged and our instrument is only sensitive to the brightest pulses. 
Therefore our detections are biased to high fluence pulses.
Among these pulsars, J1107--5907 is a known intermittent pulsar previously studied using ASKAP \citep{2016MNRAS.456.3948H}.

\begin{figure*}
\centering
\begin{tabular}{ccc}
\includegraphics[trim={0 0.0cm 0 0.0cm },clip,width=0.65\columnwidth]{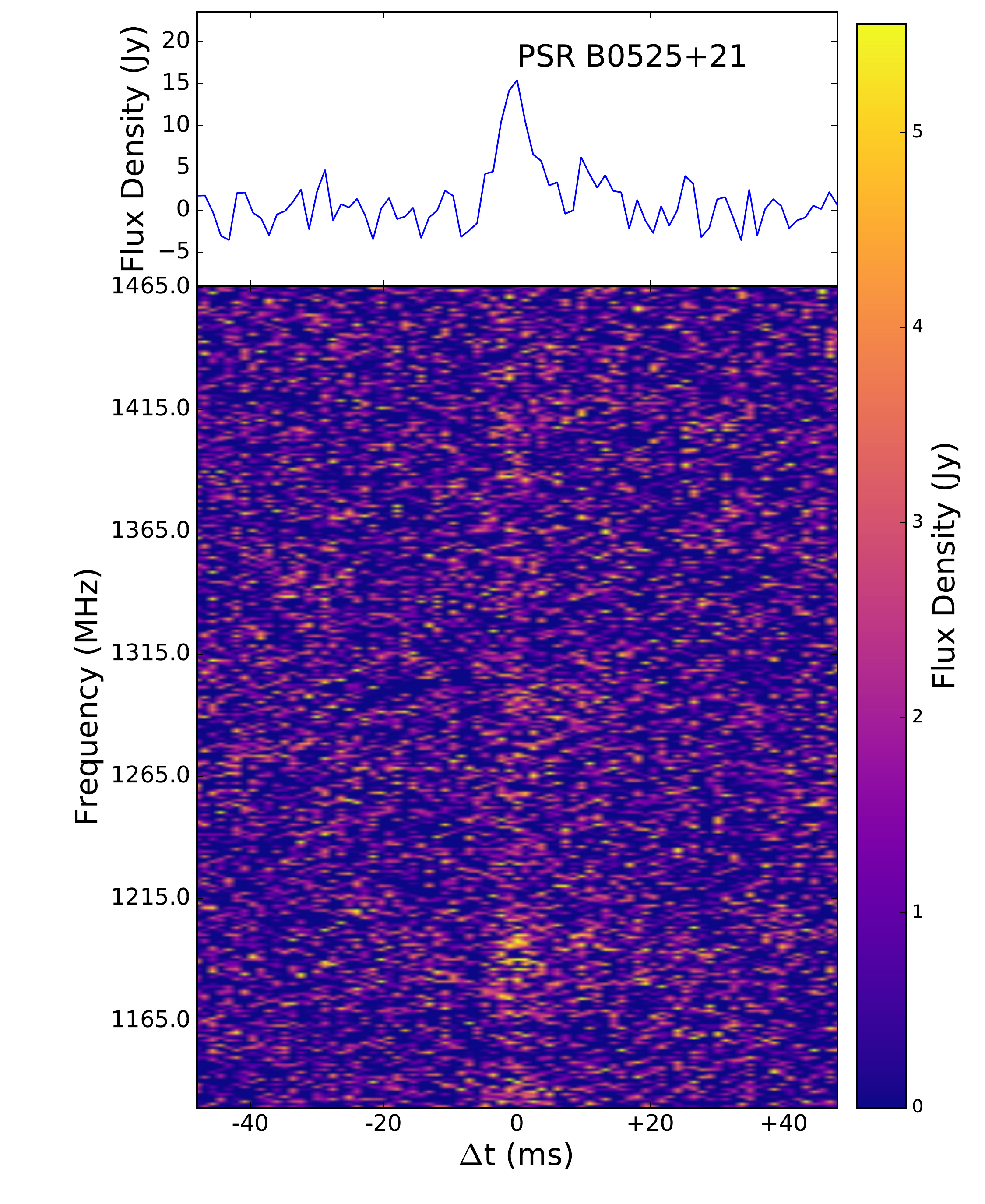} & 
\includegraphics[trim={0 0.0cm 0 0.0cm },clip,width=0.65\columnwidth]{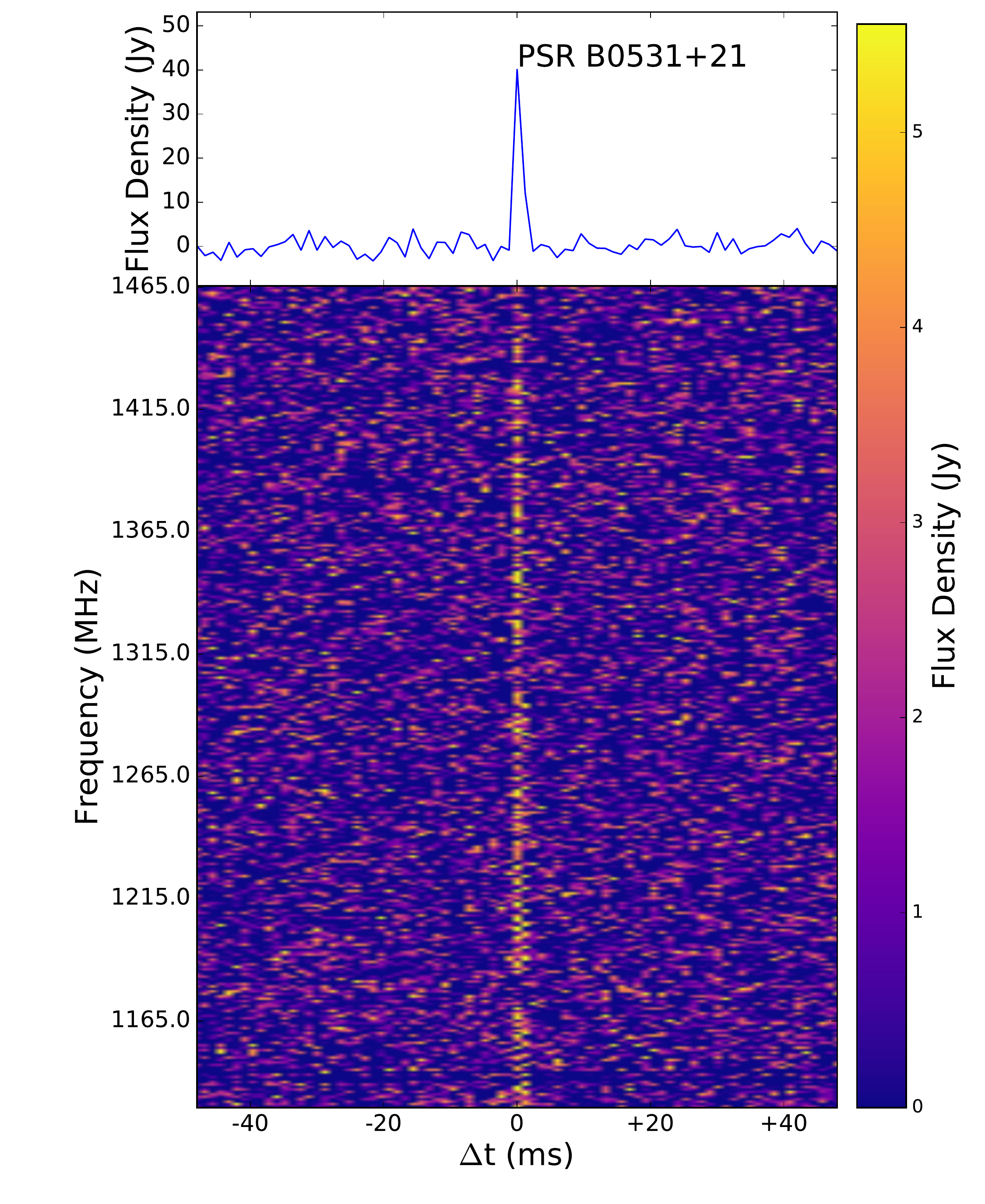}&
\includegraphics[trim={0 0.0cm 0 0.0cm },clip,width=0.65\columnwidth]{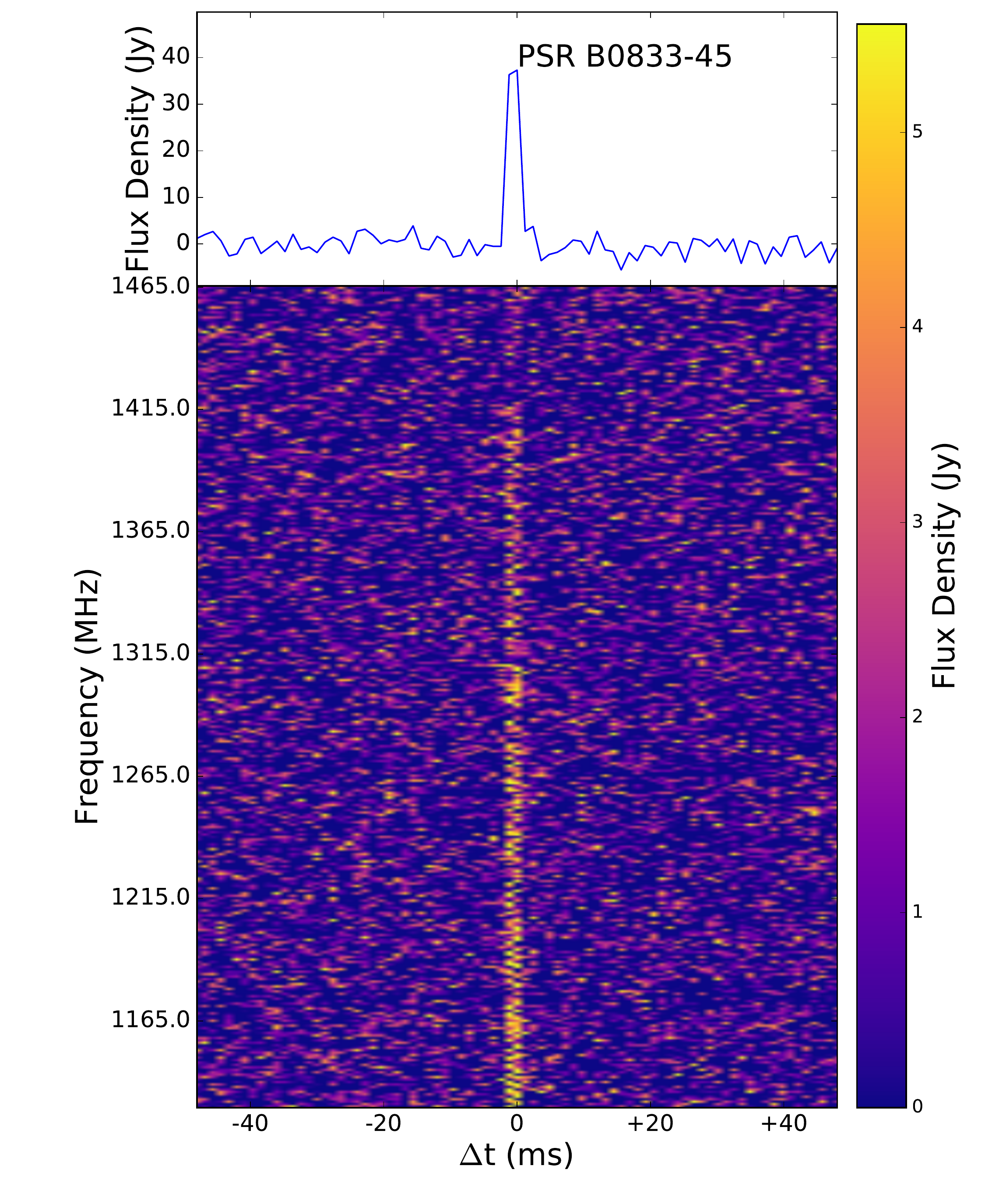}\\
\includegraphics[trim={0 0.0cm 0 0.0cm },clip,width=0.65\columnwidth]{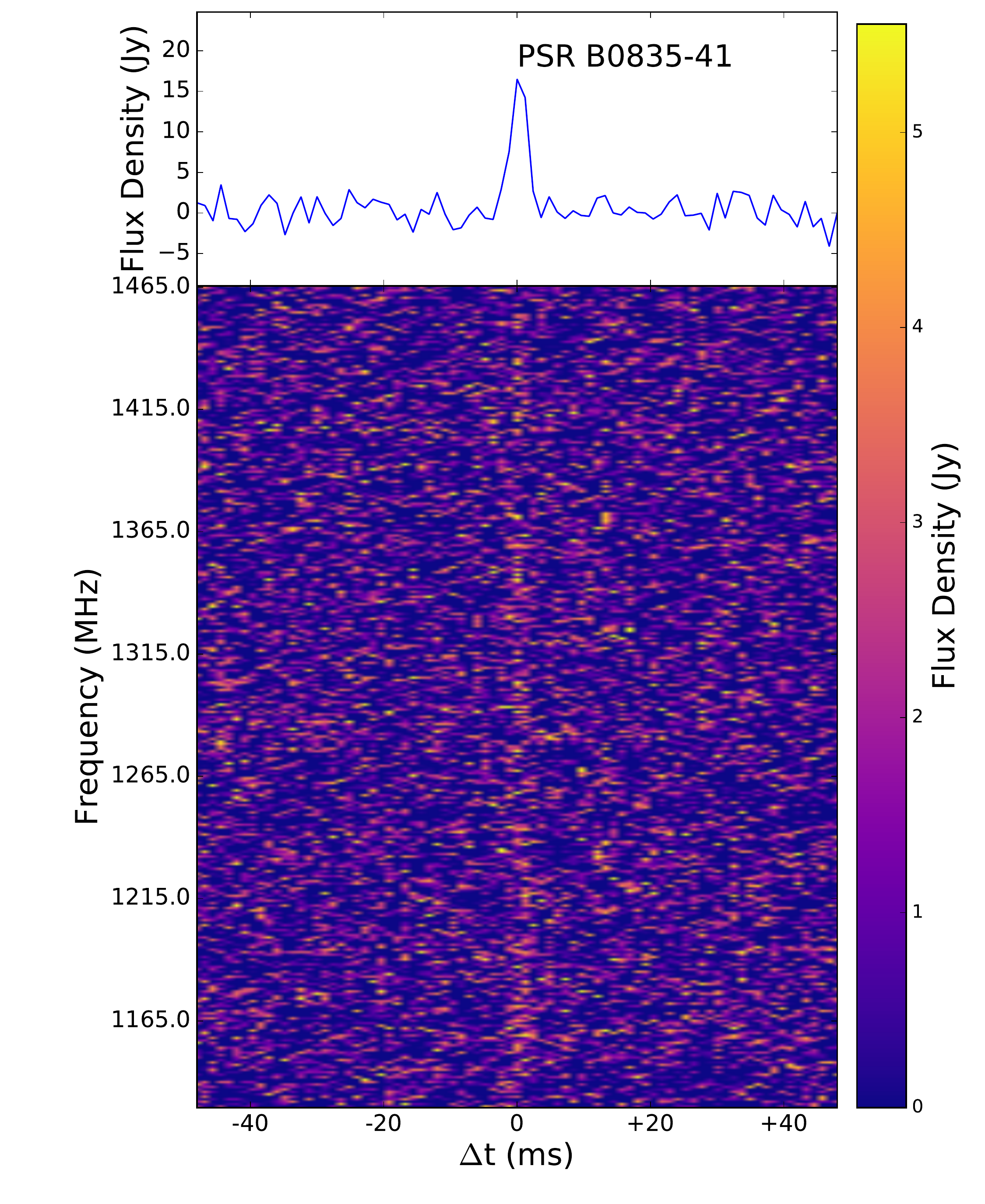}&
\includegraphics[trim={0 0.0cm 0 0.0cm },clip,width=0.65\columnwidth]{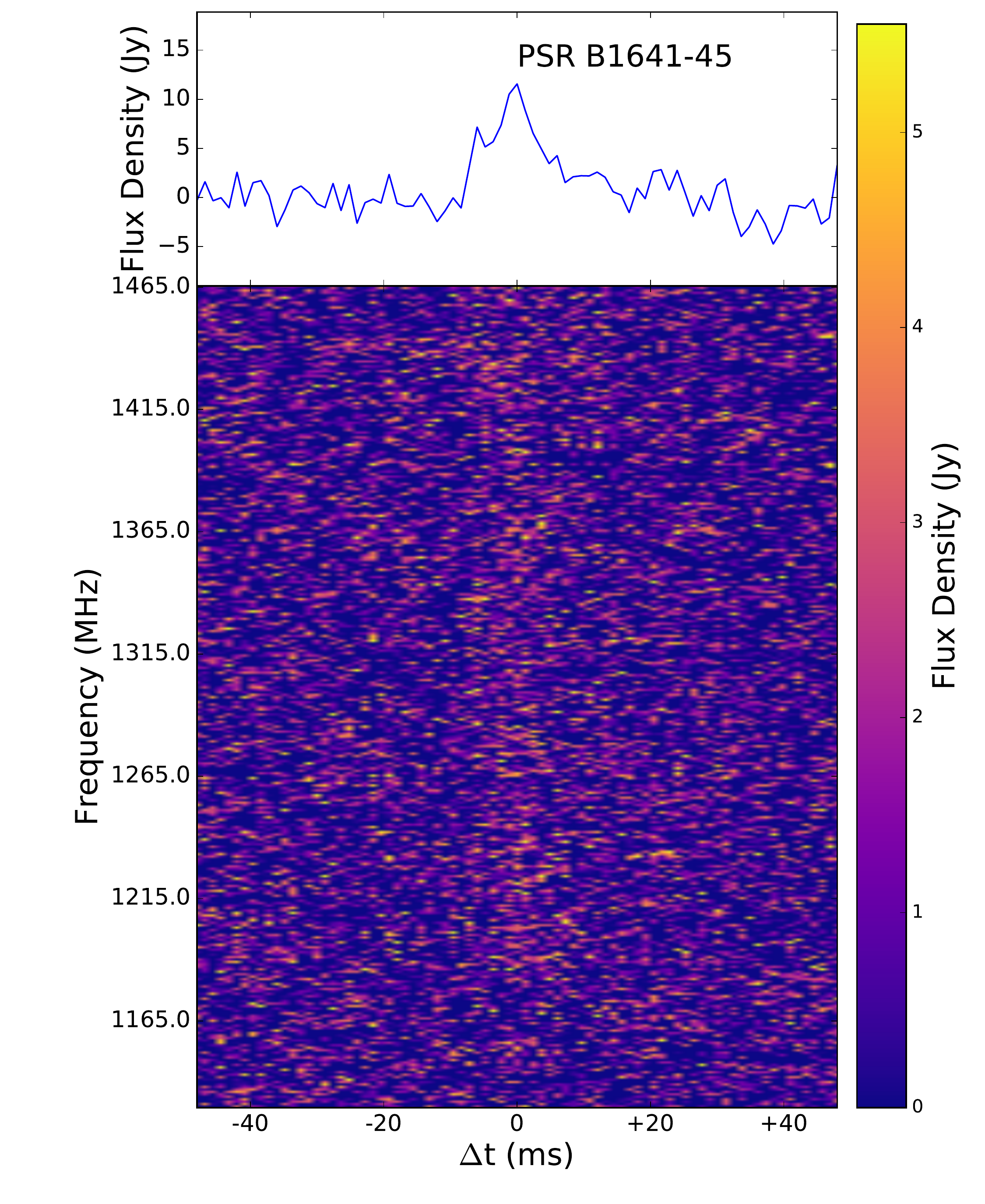}&
\includegraphics[trim={0 0.0cm 0 0.0cm },clip,width=0.65\columnwidth]{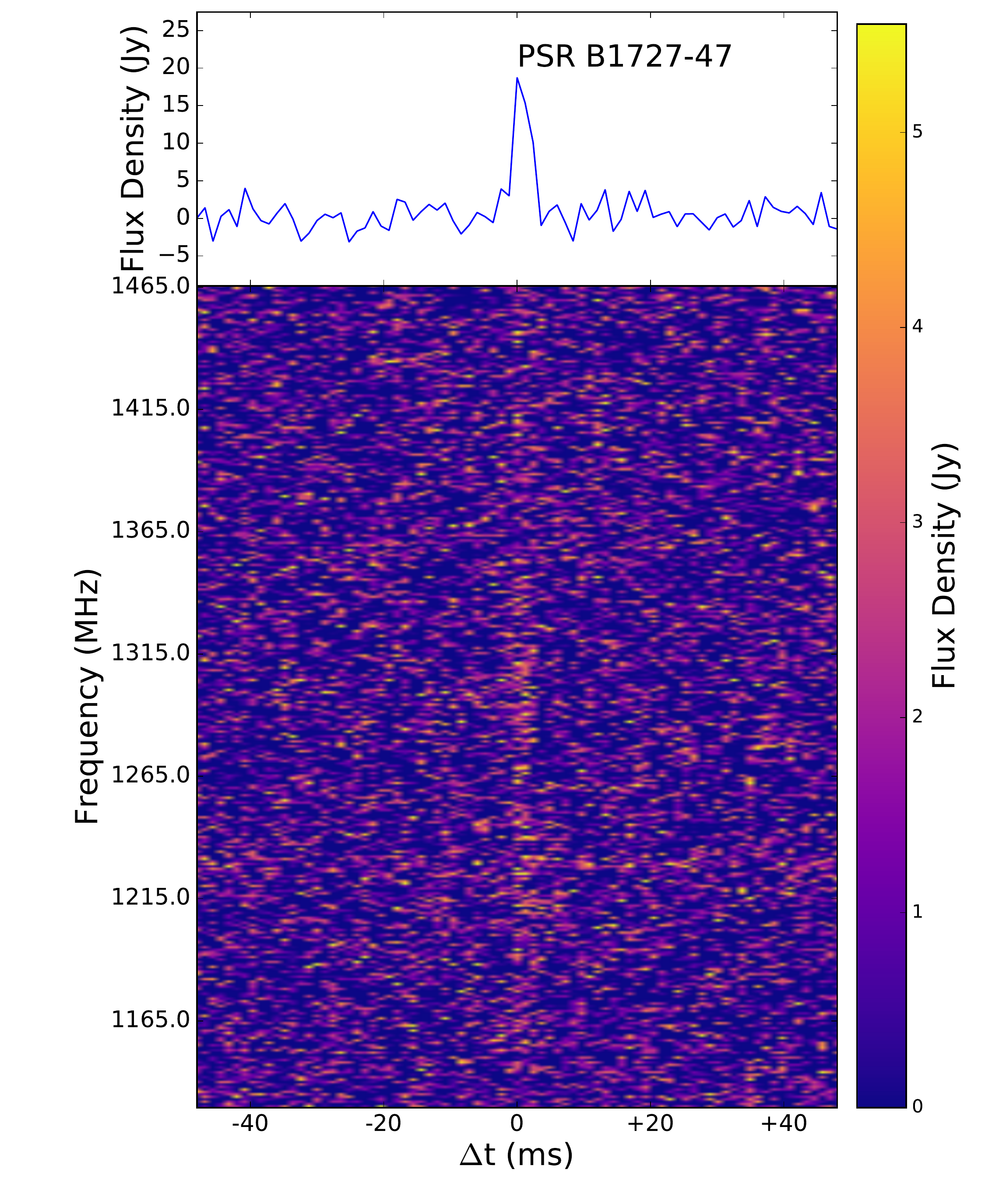}\\

\end{tabular}

\caption{Example single pulse
detections of each pulsar.}
\end{figure*}
\renewcommand{\thefigure}{\arabic{figure} (Cont.)}
\addtocounter{figure}{-1}
\begin{figure*}
\centering
\begin{tabular}{ccc}
\includegraphics[trim={0 0.0cm 0 0.0cm },clip,width=0.65\columnwidth]{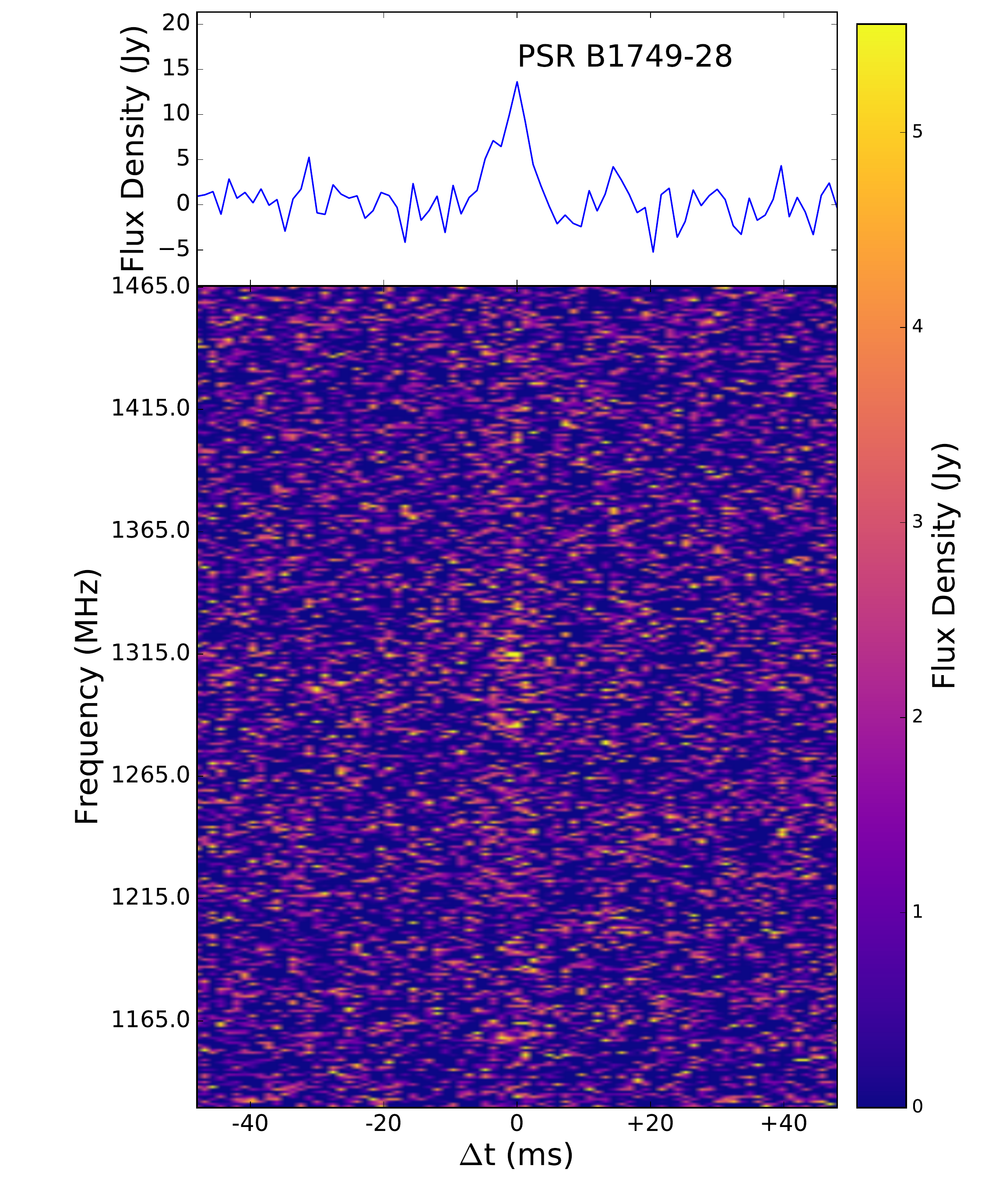}&
\includegraphics[trim={0 0.0cm 0 0.0cm },clip,width=0.65\columnwidth]{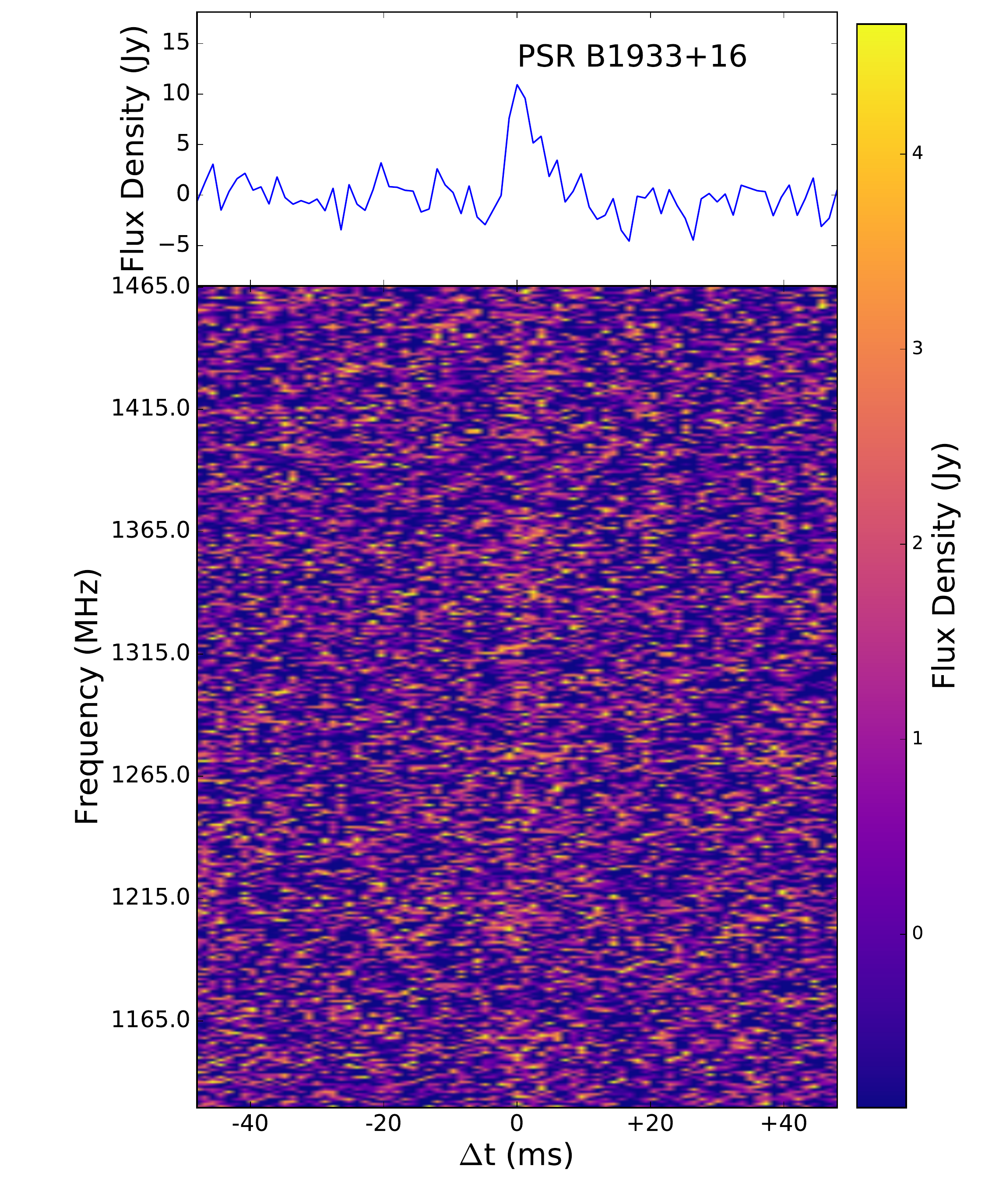}&
\includegraphics[trim={0 0.0cm 0 0.0cm },clip,width=0.65\columnwidth]{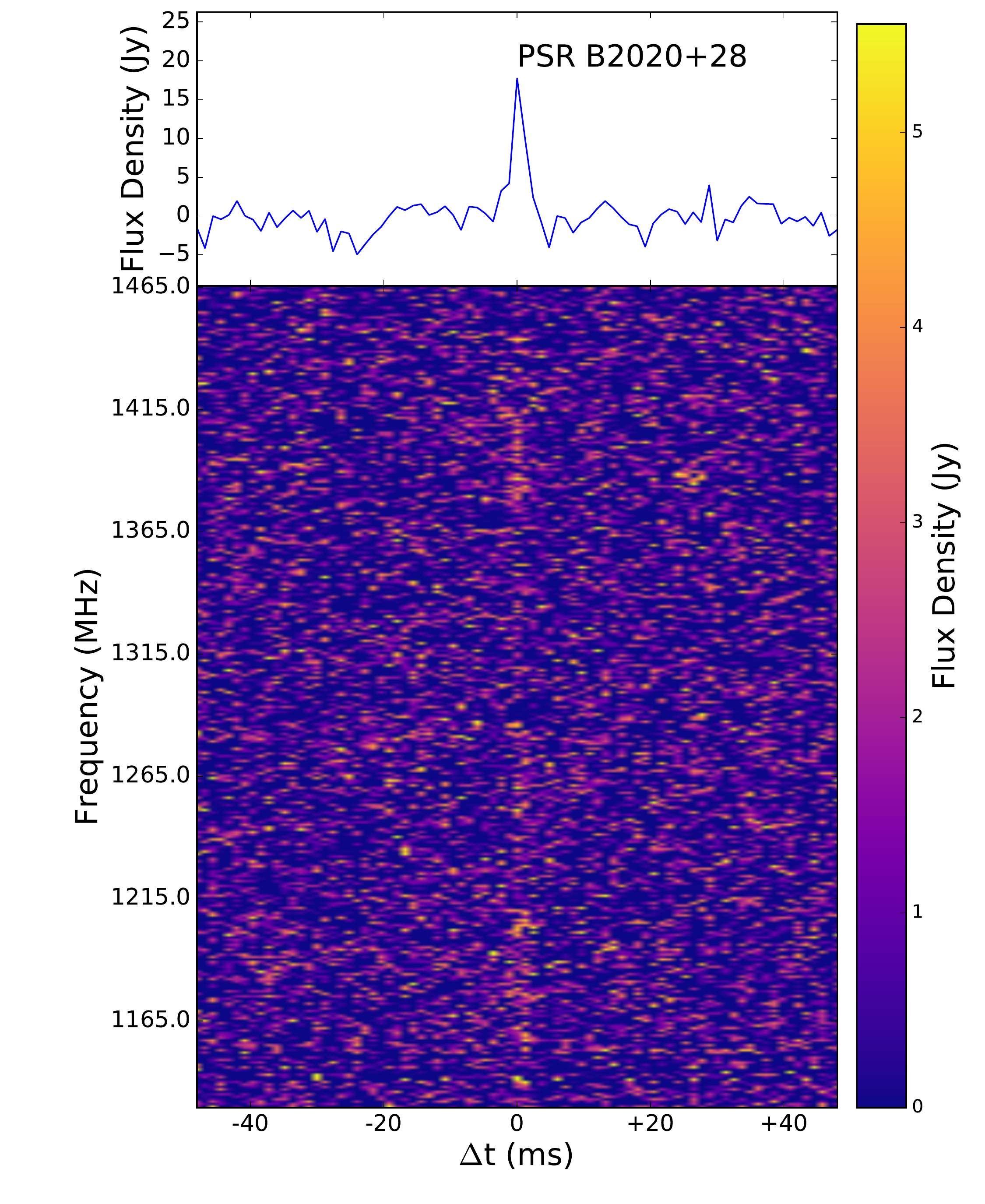}
\\
\includegraphics[trim={0 0.0cm 0 0.0cm },clip,width=0.65\columnwidth]{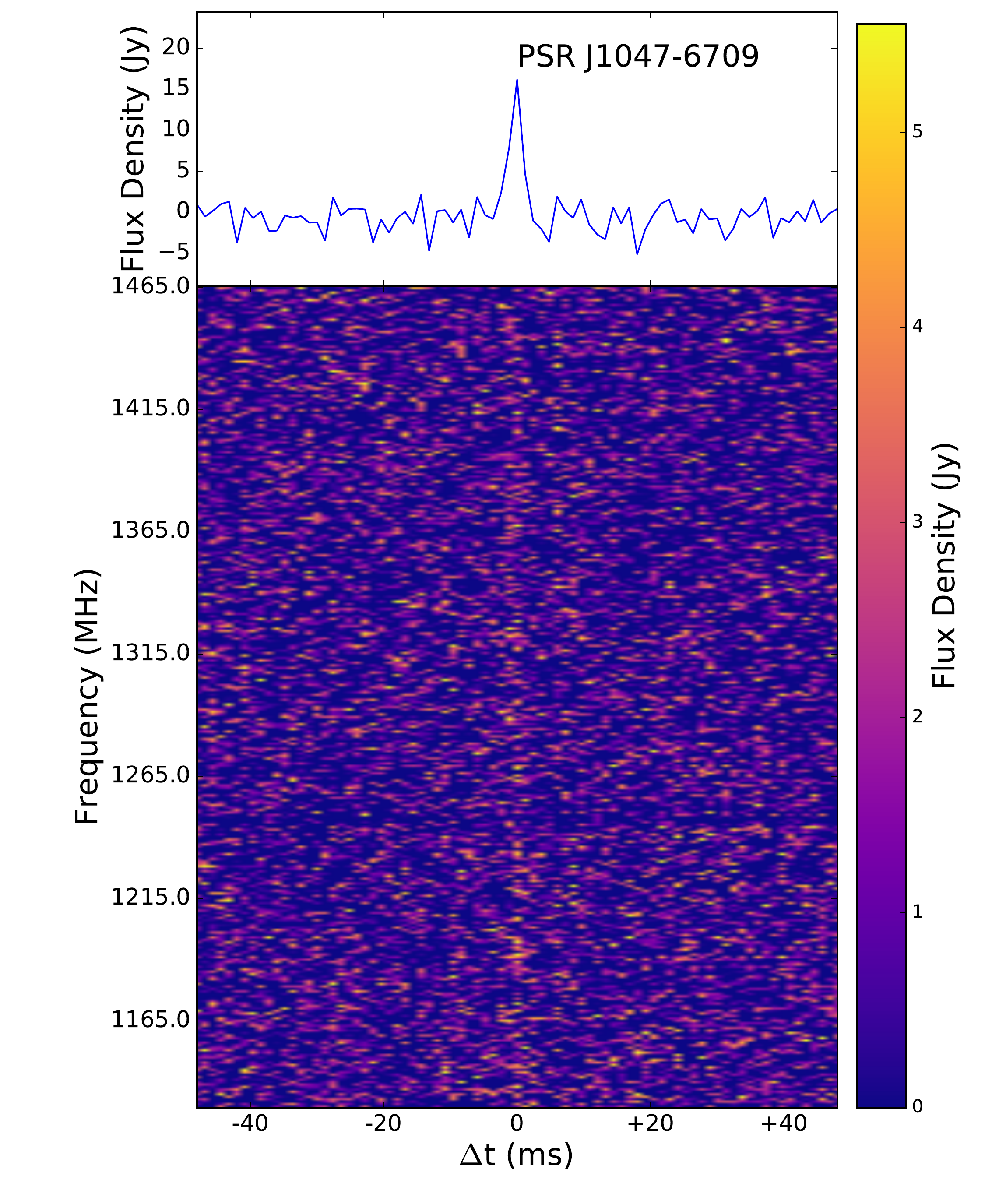}&
\includegraphics[trim={0 0.0cm 0 0.0cm },clip,width=0.65\columnwidth]{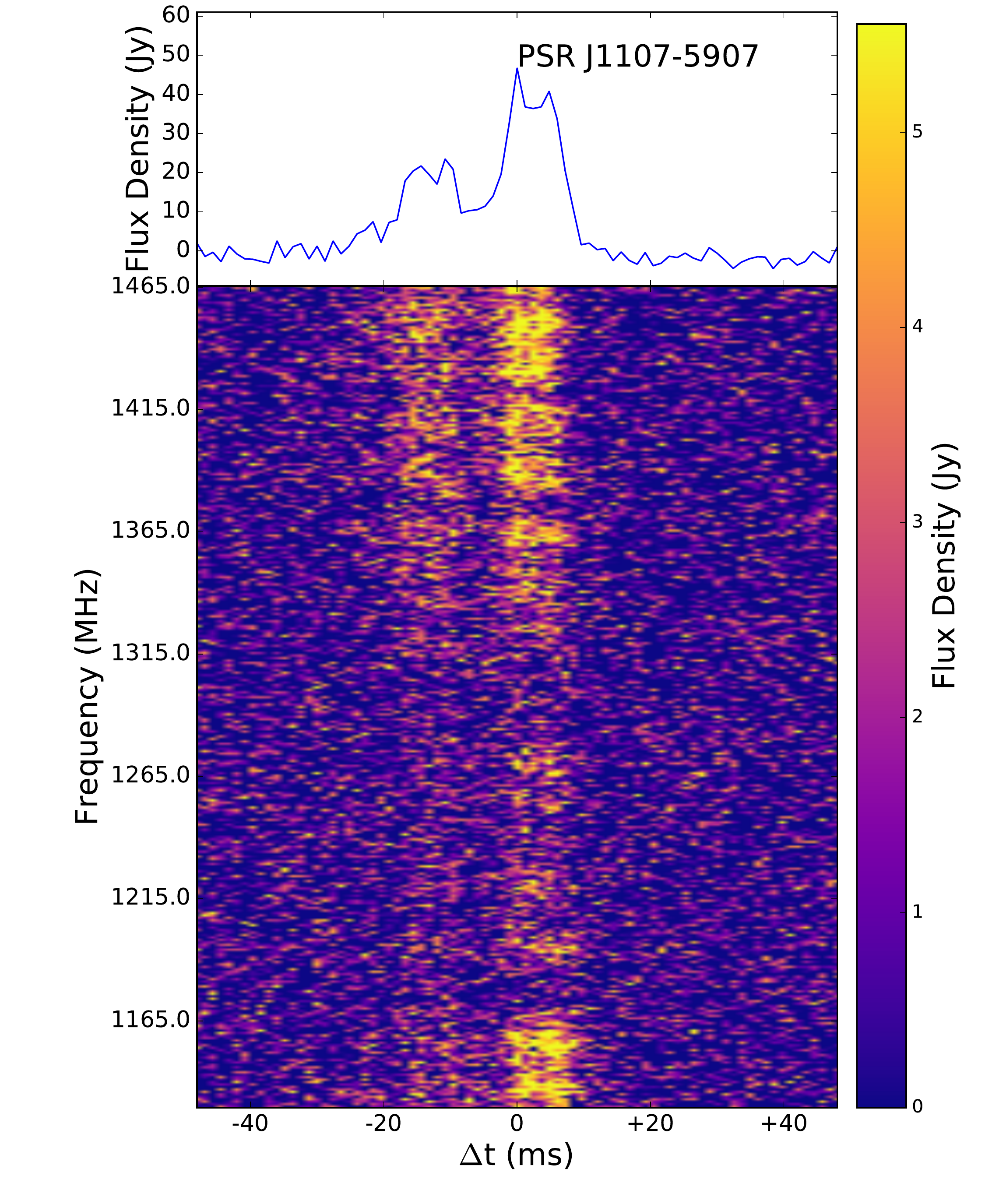}
\\
\end{tabular}
\caption{Example single pulse
detections of each pulsar.}
\end{figure*}
\renewcommand{\thefigure}{\arabic{figure}}
\subsection{Detection of FRB 180430}
We collected a total of 25079 unidentified candidates in 1514 antenna-hours. We then inspected the candidates in a time-DM plot to identify most false positive RFI events. For the remaining candidates, we checked the spectra created by the pipeline to verify the candidates and remove RFI events.
After RFI removal and visual verification, we were able to find FRB 180430.
We show the information for FRB 180430 and detection details in Tables \ref{tab:frb_dm} and \ref{tab:frb}, 
we also include the Galactic DM contribution and excess DM from both  
NE2001 \citep{2001ApJ...549..997C} and YMW16 \citep{2017ApJ...835...29Y} models for comparison.

We searched all other candidates from observations of the same pointing in a total of 10.9 hours.
We did not find any other candidates with the same dispersion measure above a S/N ratio of 7.5.
No single pulse of the same DM was detected in this pointing region during the whole survey period.
A Periodicity search of the filterbank containing FRB 180430 did not 
find any periodic signal during the observation at the given DM in a total observation time of 1675.85 seconds.
We also acquired 2.5 hours of occasional follow-up time between June 5th 2018 and December 23rd 2018 with Parkes and did not find any repetition from FRB 180430 at a fluence limit of $\sim$0.1 Jy ms.
\begin{figure}
    \centering
	\includegraphics[width=0.6\columnwidth]{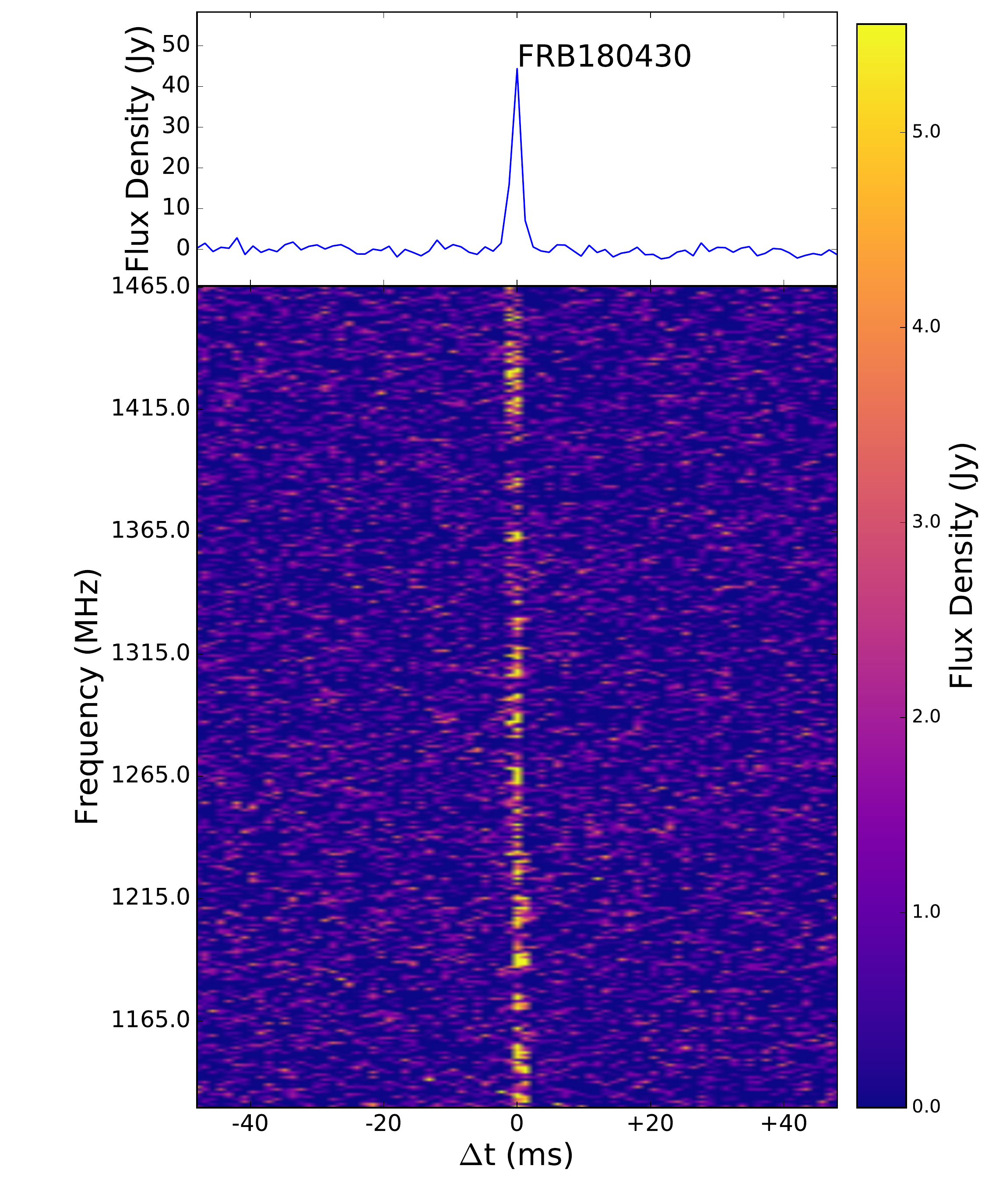}
    \caption{Average profile and dynamic spectra of FRB 180430. 
    The burst in this plot has been dedispersed to 
    the maximum-likelihood dispersion measure of 264.1 pc cm$^{-3}$. 
    The colour scale is set to range from the mean to 4$\sigma$ 
    of the off pulse intensity}
    \label{fig:frb_fb}
\end{figure}

We performed a multi-beam localisation, 
flux measurement and dispersion measurement of FRB 180430 by
combining the detection from all 4 beams.
The details of the localisation technique is explained in \citet{2017ApJ...841L..12B}. 
We are able to constrain the position of FRB 180430 
within a region of $6' \times 6'$ at 90\% 
confidence and $2' \times 2'$ at 50\% confidence.  

\begin{table}
	\centering
	\caption{Properties of FRB 180430 }
	\label{tab:frb_dm}
	\begin{tabular}{lc} 
    \hline
    \multicolumn{2}{l}{Observation Results}\\
		\hline
		Date& 2018 Apr 30\\
		Time of Arrival  & 10:00:35.70(TAI)\\
		Dispersion Measure (DM) & 264.1(5) pc cm$^{-3}$ \\
        Right Ascension$^a$ (J2000) &  $06^h51^m(6)$
 \\
        Declination$^a$ (J2000) & $-09$\degree 57$\arcmin$(6) \\
        Galactic Longitude & 221.76\degree \\
        Galactic Latitude & $-4.61$\degree \\
        Width & 1 sample$^b$\\
        Fluence & 177(4) Jy ms
 \\
\hline
\multicolumn{2}{l}{Milky Way DM Estimates}\\
		\hline
        DM$_{\text{MW,NE}2001}$ & 165.44  pc cm$^{-3}$
        \\
        DM$_{\text{MW,YMW}16}$ & 294.16  pc cm$^{-3}$
        \\
        \hline
        \hline
        \multicolumn{2}{l}{a. The 1-$\sigma$ uncertainty of the last digit is given}\\
        {in the parentheses}\\
         \multicolumn{2}{l}{b. The time of one sample in the CRAFT filterbank is 1.2 ms}\\
	\end{tabular}

\end{table}

\begin{table}
	\centering
	\caption{The observation details of the ASKAP PAF for the detection of FRB 180430.}
	\label{tab:frb}
	\begin{tabular}{lcccc} 
		\hline
		Beam & S/N & DM ($ \text{pc} \ \text{cm}^{-3}$) &
        R.A. (deg) & Declination (deg)
         \\
      
        \hline
        
           16 & 17.34 &263.4 & 103.380932 & $-10.434731$\\
           17 & 28.20 &263.4 & 102.970281 & $-9.630296$\\
           21 & 25.25&263.4 & 102.467378 & $-10.381924$\\
           22 & 14.28 &263.4 & 102.058956 & $-9.576174$\\

\hline
		\hline
	
	\end{tabular}
\end{table}

\section{Discussion}

\subsection{RRAT Non-Detections}
We previously show in our population model that we may detect 0-4 RRATs. 
In this survey, we did not detect any RRATs in our single pulse pipeline. 
It should be noted that the average pulse width of known RRAT 
pulses is 18.3 ms with a wide range from $\sim 3$ ms to $\sim $65 ms, 
the average width exceeds the $\sim 13$ ms width threshold of our current detection pipeline (see Section 2.1) which may also effects the detection. 
The average width of RRATs at 1400MHz 
as recorded in the \textsc{rratalog} is 18.28 ms,
with no clear correlation of width and fluence. 
This implies that we may not detect some single pulses 
due to the limited width threshold applied for candidate selection.

Based on the zero RRAT single pulses in a total observation of $34692.4\ \text{deg}^2\ \text{h}$,
the non-detection of RRATs provides a 95 per cent Poisson uncertainties upper limit of 
$< 8.64\E{-5}\ \text{deg}^{-2}\ \text{h}^{-1}$  within 
width $< 12.7$ ms and
above a flux threshold of 20 Jy \citep{1986ApJ...303..336G}.
This limit is heavily affected by the pipeline's candidate selection parameters, 
and we expect to perform a more complete longer width and lower DM search with improved software.

\subsection{The localisation of FRB 180430}
FRB 180430 has a relatively large DM (264.1(5) pc cm$^{-3}$) 
compared to two nearby pulsars recorded in within a 10 degree radius which all have a DM of 
$\sim 100$ pc cm$^{-3}$ this makes it very hard to estimate the DM contribution of the Milky Way in this direction.
The distance of these pulsars based on their dispersion measurement are $\sim 2$ kpc.
Both pulsars have much lower dispersion than FRB 180430 and there is a lack of pulsars
at higher dispersion measures in this direction.
There is a disagreement on the Galactic DM contribution of FRB 180430 from the two main Galactic electron density models.
It is not significantly higher than the estimate DM from the models as shown in Table \ref{tab:frb_dm}.

This is the second FRB, the other being FRB 010621, that has a disputable DM estimate, we try and explain which model is more feasible.
The measurement from NE2001 indicates 
an excess DM of 98.7 $ \text{pc} \ \text{cm}^{-3}$, suggesting an extragalactic object.
The YMW16 model provides a higher DM which exceeds the DM of FRB 180430, indicating a Galactic object.
However the YMW Milky Way model arbitrarily extends the disk beyond
$\sim$19 kpc in the anticenter direction due to the lack of pulsars in that direction 
beyond 15 kpc \citep{2017ApJ...835...29Y}. 
This may result in an overestimated Galactic DM from the YMW16 model.
In this case, we suggest the lower Galactic DM by NE2001 would be more correct.

We consider a 12 $ \text{pc} \ \text{cm}^{-3}$ 
contribution from the Milky Way halo (DM$_{\rm Halo}$) following \citet{2018ApJ...867L..10M}, using the excess DM of pulsars detected on the near side of the Large Magellanic Cloud \citep{2005AJ....129.1993M},
The halo contribution is uncertain and may be higher with additional DM at further distances giving DM$_{\rm Halo}$ = 30  $ \text{pc} \ \text{cm}^{-3}$
\citep{2018ApJ...867L..10M,2015MNRAS.451.4277D}, we use the former lower estimate to determine maximum redshift.

The DM of radio single pulses usually provides an estimate for the distance of these phenomena.
In this case,
after subtracting the NE2001 model and halo DM contribution, the estimated maximum excess extragalactic DM is 86.7 $ \text{pc} \ \text{cm}^{-3}$.
This gives an estimate a maximum host galaxy redshift (zero host DM contribution) of
$z\approx 0.072$ \citep{2004MNRAS.348..999I}. 
However, we note that there is a large scatter effect in such low dispersion measurements for estimating the redshift \citep{2014ApJ...780L..33M}.

We could not find any counterpart object such as a X-ray/radio pulsar, supernova remnant or 
host galaxy in {SIMBAD} \citep{2000A&AS..143....9W}, the High Energy Astrophysics Science Archive Research Center (HEASARC)\footnote{https://heasarc.gsfc.nasa.gov/} and the NASA/IPAC Extragalactic Database (NED)\footnote{https://ned.ipac.caltech.edu/} using the current $6' \times 6'$ localisation information of FRB 180430. 
The lack of host galaxies in the search results is expected due to the Galactic plane is not often observed in galaxy surveys.

\subsection{Is FRB 180430 a RRAT?}
The low excess DM of FRB 180430 and disagreement with the 
YMW16 model does not allow us to exclude the possibility of this FRB being a galactic pulsar giant pulse or RRAT outburst. 
The excess DM in these cases clearly cannot distinguish an FRB and a RRAT, further analysis on the repetition rate and
 pulse properties such as scattering
 would help the argument for either case \citep{2016MNRAS.459.1360K}.
We take FRB 180430 as an example case to try discuss the boundaries between FRBs and RRATs at the edge of the Milky Way.

The absence of a supernova remnant, X-ray source or radio pulsar does not favour the 
possibility of this radio burst originating from a Galactic pulsar.
However this does not exclude the possibility of a very faint source of Galactic origin for the burst.
We note that there were no X-ray observations for this field archived in HEASARC. Future observations in X-ray may be able to further reveal the nature of FRB 180430.

The narrow width and patchy non-scattering spectral 
properties of FRB 180430 are similar to other FRBs 
detected by ASKAP  and is narrower
compared to RRATs which are usually wider in pulse width.
We searched for possible H$\alpha$ and H$\beta$ excess in the SuperCOSMOS \citep{2005MNRAS.362..689P} 
and SHASSA \citep{2001PASP..113.1326G} but
did not find any detection within the localisation region 
to compare with the work performed in \citet{2014MNRAS.440..353B}.
There is also no scattering observed in FRB 180430, which would likely suggest less ISM contribution in the total medium FRB 180430 has passed through \citep{2013MNRAS.436L...5L}. These facts favour towards the NE2001 model and suggest an extragalactic origin for FRB180430.

If FRB 180430 originated in the Milky Way, 
it could be possible that it was a giant pulse from
a young pulsar in a supernova remnant (SNR) 
similar to PSR B0531+21  \citep[Crab pulsar,][]{2016MNRAS.458L..19C,2016MNRAS.462..941L} or PSR B0540$-$69 \citep{1984ApJ...287L..19S,1993ApJ...403L..29M,2004MNRAS.355...31J} in the Large Magellanic Cloud.
There are observations of extremely narrow mega-Jansky level Crab giant pulses \citep{2007ApJ...670..693H}.
Based on the YMW16 Galactic model, a pulsar with the DM of FRB 180430 would be at an estimate distance of no less than 8 kpc.
The crab pulsar is at a distance of 2 kpc, the observed flux of a giant pulse similar to that of the Crab pulsar at such distance would be $\sim 60$ Jy. 
This is to the same order of flux density to
FRB 180430 and suggests the possibility of a pulsar giant pulse at
the edge of the Galactic plane. 
We performed a periodicity search with 1675.85 seconds of data from this
observation.
We did not find any periodicity at the DM of the FRB above $3\sigma$ (approximately 0.06 Jy with an assumed duty cycle of 0.1).
These giant pulses should frequently appear, if it is indeed a pulsar we expect to see repetition from this source.
However in the 10 hours of observation with ASKAP and 2.5 hours follow up with Parkes, we have not yet seen any repetition from this region.

We also consider the possibility of a RRAT at such distances.
Although there has been X-ray emission detected from RRATs such as J1819-1458 \citep{2006ApJ...639L..71R},  
there are also cases where no X-ray emission has been detected \citep{2009MNRAS.400.1445K}. Therefore X-ray emission may not play an important role in identifying FRB 180430 as a RRAT.
Most RRATs have a pulse width of $\sim15$ ms and thus are much wider than FRB 180430 and other FRBs, but J1819-1458 is actually relatively narrow and has a width of 3.6 ms. 
J1819-1458 is also the brightest RRAT observed, which peak flux reaches 3.6 Jy is at a distance of 3.6 kpc
\citep{2006Natur.439..817M}.
Should FRB 180430 be a RRAT, it would be the brightest RRAT outburst and the narrowest pulse profile observed, yet these properties fit very well to the generic appearance of the FRB population. 
Also FRB 180430 has yet been observed to repeat as mentioned previously, such non-repetition would not be favoured towards RRATs
\citep{2016MNRAS.459.1360K}.

In summary, 
despite the possibilities of RRATs and pulsar giant pulses no other giant pulse of the same DM has been detected yet from this position.
The no repetition of a high fluence unscattered single pulse event suggests that FRB 180430 is an extragalactic radio burst compared to the other possibility of a young pulsar or RRAT in the Milky Way.


\subsection{FRB event rate and single pulse limits}
Based on the total search time and number of FRB detections, 
we provide an estimate comparison FRB rate from this Galactic plane survey.
The Galactic plane survey consisted 160 pointings, each pointing's PAF footprint had an effective 
FoV of 22.9 deg$^2$. We use this effective FoV to calculate the total exposure of the survey. 
The resulting event rate
\begin{equation}                             {\cal R} =                                   
    1\ \text{FRBs}\times 
    \frac{24\ \text{h}\ \text{d}^{-1}\times41253\ \text{deg}^2\ \text{sky}^{-1}}
    {34692.4\ \text{deg}^2\ \text{h}}
	\label{eq:frb_rate}
\end{equation}
which translates to ${\cal R} = 29^{+107}_{-27}\ \text{sky}^{-1}\ \text{d}^{-1}$ 
with 95 per cent Poisson uncertainties \citep{1986ApJ...303..336G} at 20 Jy ms fluence equivalent sensitivity.
The event rate, albeit with large uncertainties due to one single detection, is comparable to the event rate at Galactic latitude $50 \degree$
presented in  \citet{2018Natur.562..386S}.
%

\section{Conclusion}
In this paper, we present the results of a 10 hour per pointing fly's-eye survey of the Galactic plane with ASKAP in a total observation time of 63.117 antenna-days on the Galactic plane between $|b|< 7\degree$.
We then present the data analysis and results in search of fast radio bursts and other fast radio transients.
The main results and conclusions of this survey are as follows:

\begin{enumerate}
\item{No single pulse from RRATs were detected during this survey.
Non-detection of RRATs gives an upper limit of the RRAT detection rate  $< 8.64\E{-5} \text{deg}^{-2}\ \text{h}^{-1}$ at 20 Jy ms fluence equivalent sensitivity and width < 12.7 ms. }
\item{
We detected single pulses from 12 known pulsars.}

\item{FRB 180430 is an anti-centre FRB candidate with an uncertain low extragalactic excess DM between 
zero and {98.7 $ \text{pc} \ \text{cm}^{-3}$}, we suggest this is more likely an extragalactic source a possible host galaxy at redshift $z< 0.072$.
No catalogued galaxy was found and follow up with the Parkes radio telescope did not detect any repetition.}
\item{The detection of one FRB during this survey gives 
a FRB detection rate of $29^{+107}_{-27}\ \text{sky}^{-1}\ \text{d}^{-1}$ at 20 Jy ms fluence equivalent sensitivity. 
}

\end{enumerate}

\section*{Acknowledgements}

The Australian SKA Pathfinder and the Parkes radio telescope is part of the Australia
Telescope National Facility which is managed by CSIRO.
Operation of ASKAP is funded by the Australian Government
with support from the National Collaborative Research
Infrastructure Strategy. ASKAP uses the resources of the
Pawsey Supercomputing Centre. Establishment of ASKAP, the
Murchison Radio-astronomy Observatory and the Pawsey
Supercomputing Centre are initiatives of the Australian
Government, with support from the Government of Western
Australia and the Science and Industry Endowment Fund. We
acknowledge the Wajarri Yamatji people as the traditional
owners of the Observatory site.
HQ acknowledges the support of 
a Hunstead Merit Award for Astrophysics from 
the Hunstead Gift for Astrophysics.
KWB and RMS acknowledge the support of the Australian Research Council through grant DP18010085.  
RMS acknowledges salary support from the ARC through grants FL150100148 and  CE17010004.
TM acknowledges the support of the Australian Research Council through grant FT150100099
This research has made use of the NASA/IPAC Extragalactic Database (NED), which is operated by the Jet Propulsion Laboratory, California Institute of Technology, under contract with the National Aeronautics and Space Administration.
This research has made use of the SIMBAD database,
operated at CDS, Strasbourg, France.
DRL and DA
acknowledge support from the Research Corporation for Scientific Advancement and the National Science Foundation awards AAG-1616042, OIA-1458952 and PHY-1430284.



\bibliographystyle{mnras}
\bibliography{export-bibtex} 




\bsp	
\label{lastpage}
\end{document}